\documentclass[twocolumn]{aastex631}

\newcommand{\M}{M$_\odot$}
\newcommand{\ergs}{erg\,s$^{-1}$}
\newcommand{\kms}{km\,s$^{-1}$}

\shorttitle{LFCs in red galaxies}
\shortauthors{Nicholl et al.}

\begin{document}

\title{AT2022aedm and a new class of luminous, fast-cooling transients in elliptical galaxies}



\author[0000-0002-2555-3192]{M. Nicholl}
\affiliation{Astrophysics Research Centre, School of Mathematics and Physics, Queens University Belfast, Belfast BT7 1NN, UK}\email{matt.nicholl@qub.ac.uk}

\author[0000-0003-4524-6883]{S. Srivastav}
\affiliation{Astrophysics Research Centre, School of Mathematics and Physics, Queens University Belfast, Belfast BT7 1NN, UK}

\author{M. D. Fulton}
\affiliation{Astrophysics Research Centre, School of Mathematics and Physics, Queens University Belfast, Belfast BT7 1NN, UK}

\author[0000-0001-6395-6702]{S. Gomez}
\affiliation{Space Telescope Science Institute (STScI), 3700 San Martin Drive, Baltimore, MD 21218, USA}

\author{M. E. Huber}
\affiliation{Institute for Astronomy, University of Hawaii, 2680 Woodlawn Drive, Honolulu HI 96822, USA}

\author{S. R. Oates}
\affiliation{School of Physics and Astronomy, University of Birmingham, Birmingham B15 2TT, UK}
\affiliation{Institute for Gravitational Wave Astronomy, University of Birmingham, Birmingham B15 2TT, UK}

\author[0009-0009-2627-2884]{P. Ramsden}
\affiliation{School of Physics and Astronomy, University of Birmingham, Birmingham B15 2TT, UK}
\affiliation{Institute for Gravitational Wave Astronomy, University of Birmingham, Birmingham B15 2TT, UK}

\author[0000-0003-2705-4941]{L. Rhodes}
\affiliation{Department of Physics, University of Oxford, Denys Wilkinson Building, Keble Road, Oxford OX1 3RH, UK}

\author[0000-0002-8229-1731]{S. J. Smartt}
\affiliation{Department of Physics, University of Oxford, Denys Wilkinson Building, Keble Road, Oxford OX1 3RH, UK}
\affiliation{Astrophysics Research Centre, School of Mathematics and Physics, Queens University Belfast, Belfast BT7 1NN, UK}

\author[0000-0001-9535-3199]{K. W. Smith}
\affiliation{Astrophysics Research Centre, School of Mathematics and Physics, Queens University Belfast, Belfast BT7 1NN, UK}

\author[0000-0002-9085-8187]{A. Aamer}
\affiliation{School of Physics and Astronomy, University of Birmingham, Birmingham B15 2TT, UK}
\affiliation{Institute for Gravitational Wave Astronomy, University of Birmingham, Birmingham B15 2TT, UK}
\affiliation{Astrophysics Research Centre, School of Mathematics and Physics, Queens University Belfast, Belfast BT7 1NN, UK}

\author{J. P. Anderson}
\affiliation{European Southern Observatory, Alonso de C\'ordova 3107, Casilla 19, Santiago, Chile}
\affiliation{Millennium Institute of Astrophysics MAS, Nuncio Monsenor Sotero Sanz 100, Off. 104, Providencia, Santiago, Chile}

\author[0000-0002-8686-8737]{F. E. Bauer}
\affiliation{Instituto de Astrof\'{ı}sica, Facultad de F\'{ı}sica, Pontificia Universidad Cat\'{o}lica de Chile, Campus San Joaqu\'{ı}n, Av. Vicu\~{n}a Mackenna 4860, Macul Santiago, Chile, 7820436}
\affiliation{Centro de Astroingenier\'{ı}a, Facultad de F\'{ı}sica, Pontificia Universidad Cat\'{o}lica de Chile, Campus San Joaqu\'{ı}n, Av. Vicu\~{n}a Mackenna 4860, Macul Santiago, Chile, 7820436}
\affiliation{Millennium Institute of Astrophysics MAS, Nuncio Monsenor Sotero Sanz 100, Off. 104, Providencia, Santiago, Chile}

\author{E. Berger}
\affiliation{Center for Astrophysics $\vert$ Harvard \& Smithsonian, Cambridge, MA 02138, USA}

\author{T. de Boer}
\affiliation{Institute for Astronomy, University of Hawaii, 2680 Woodlawn Drive, Honolulu HI 96822, USA}

\author[0000-0001-6965-7789]{K. C. Chambers}
\affiliation{Institute for Astronomy, University of Hawaii, 2680 Woodlawn Drive, Honolulu HI 96822, USA}

\author[0000-0002-0326-6715]{P. Charalampopoulos}
\affiliation{Department of Physics and Astronomy, University of Turku, Vesilinnantie 5, FI-20500, Finland}

\author[0000-0002-1066-6098]{T.-W. Chen}
\affiliation{Technische Universit{\"a}t M{\"u}nchen, TUM School of Natural Sciences, Physik-Department, James-Franck-Stra{\ss}e 1, 85748 Garching, Germany}

\author{R. P. Fender}
\affiliation{Department of Physics, University of Oxford, Denys Wilkinson Building, Keble Road, Oxford OX1 3RH, UK}

\author{M. Fraser}
\affiliation{School of Physics, O'Brien Centre for Science North, University College Dublin, Belfield, Dublin 4, Ireland}

\author[0000-0003-1015-5367]{H. Gao}
\affiliation{Institute for Astronomy, University of Hawaii, 2680 Woodlawn Drive, Honolulu HI 96822, USA}

\author[0000-0003-3189-9998]{D. A. Green}
\affiliation{Astrophysics Group, Cavendish Laboratory, 19 J. J. Thomson Avenue, Cambridge CB3 0HE}

\author[0000-0002-1296-6887]{L. Galbany}
\affiliation{Institute of Space Sciences (ICE-CSIC), Campus UAB, Carrer de Can Magrans, s/n, E-08193 Barcelona, Spain}
\affiliation{Institut d'Estudis Espacials de Catalunya (IEEC), E-08034 Barcelona, Spain}

\author{B. P. Gompertz}
\affiliation{School of Physics and Astronomy, University of Birmingham, Birmingham B15 2TT, UK}
\affiliation{Institute for Gravitational Wave Astronomy, University of Birmingham, Birmingham B15 2TT, UK}

\author[0000-0002-1650-1518]{M. Gromadzki}
\affiliation{Astronomical Observatory, University of Warsaw, Al. Ujazdowskie 4, 00-478 Warszawa, Poland}

\author[0000-0003-2375-2064]{C. P. Guti\'errez}
\affiliation{Institut d'Estudis Espacials de Catalunya (IEEC), E-08034 Barcelona, Spain}
\affiliation{Institute of Space Sciences (ICE-CSIC), Campus UAB, Carrer de Can Magrans, s/n, E-08193 Barcelona, Spain}

\author{D. A. Howell}
\affiliation{Las Cumbres Observatory, 6740 Cortona Drive, Suite 102, Goleta, CA 93117-5575, USA}
\affiliation{Department of Physics, University of California, Santa Barbara, CA 93106-9530, USA}

\author[0000-0002-3968-4409]{C. Inserra}
\affiliation{Cardiff Hub for Astrophysics Research and Technology, School of Physics \& Astronomy, Cardiff University, Queens Buildings, The Parade, Cardiff, CF24 3AA, UK}

\author[0000-0001-5679-0695]{P. G. Jonker}
\affiliation{Department of Astrophysics/IMAPP, Radboud University, PO Box 9010, 6500 GL Nijmegen, The Netherlands}
\affiliation{SRON, Netherlands Institute for Space Research, Niels Bohrweg 4, 2333 CA Leiden, the Netherlands}

\author{M. Kopsacheili}
\affiliation{Institut d'Estudis Espacials de Catalunya (IEEC), E-08034 Barcelona, Spain}
\affiliation{Institute of Space Sciences (ICE-CSIC), Campus UAB, Carrer de Can Magrans, s/n, E-08193 Barcelona, Spain}

\author{T. B. Lowe}
\affiliation{Institute for Astronomy, University of Hawaii, 2680 Woodlawn Drive, Honolulu HI 96822, USA}

\author[0000-0002-7965-2815]{E. A. Magnier}
\affiliation{Institute for Astronomy, University of Hawaii, 2680 Woodlawn Drive, Honolulu HI 96822, USA}

\author{C. McCully}
\affiliation{Las Cumbres Observatory, 6740 Cortona Drive, Suite 102, Goleta, CA 93117-5575, USA}

\author[0000-0003-3255-3139]{S. L. McGee}
\affiliation{School of Physics and Astronomy, University of Birmingham, Birmingham B15 2TT, UK}
\affiliation{Institute for Gravitational Wave Astronomy, University of Birmingham, Birmingham B15 2TT, UK}

\author[0000-0001-8385-3727]{T. Moore}
\affiliation{Astrophysics Research Centre, School of Mathematics and Physics, Queens University Belfast, Belfast BT7 1NN, UK}

\author[0000-0003-3939-7167]{T. E. M\"uller-Bravo}
\affiliation{Institute of Space Sciences (ICE-CSIC), Campus UAB, Carrer de Can Magrans, s/n, E-08193 Barcelona, Spain}
\affiliation{Institut d'Estudis Espacials de Catalunya (IEEC), E-08034 Barcelona, Spain}

\author{M. Newsome}
\affiliation{Las Cumbres Observatory, 6740 Cortona Drive, Suite 102, Goleta, CA 93117-5575, USA}
\affiliation{Department of Physics, University of California, Santa Barbara, CA 93106-9530, USA}

\author{E. Padilla Gonzalez}
\affiliation{Las Cumbres Observatory, 6740 Cortona Drive, Suite 102, Goleta, CA 93117-5575, USA}
\affiliation{Department of Physics, University of California, Santa Barbara, CA 93106-9530, USA}

\author[0000-0002-7472-1279]{C. Pellegrino}
\affiliation{Las Cumbres Observatory, 6740 Cortona Drive, Suite 102, Goleta, CA 93117-5575, USA}
\affiliation{Department of Physics, University of California, Santa Barbara, CA 93106-9530, USA}

\author[0000-0001-6540-0767]{T. Pessi}
\affiliation{Instituto de Estudios Astrof\'isicos, Facultad de Ingenier\'ia y Ciencias, Universidad Diego Portales, Av. Ej\'ercito Libertador 441, Santiago, Chile}

\author[0000-0003-4663-4300]{M. Pursiainen}
\affiliation{DTU Space, National Space Institute, Technical University of Denmark, Elektrovej 327, 2800 Kgs. Lyngby, Denmark}

\author[0000-0002-4410-5387]{A. Rest}
\affiliation{Space Telescope Science Institute, Baltimore, MD 21218, USA}
\affiliation{Department of Physics and Astronomy, The Johns Hopkins University, Baltimore, MD 21218, USA}

\author{E. J. Ridley}
\affiliation{School of Physics and Astronomy, University of Birmingham, Birmingham B15 2TT, UK}
\affiliation{Institute for Gravitational Wave Astronomy, University of Birmingham, Birmingham B15 2TT, UK}

\author{B. J. Shappee}
\affiliation{Institute for Astronomy, University of Hawaii, 2680 Woodlawn Drive, Honolulu HI 96822, USA}

\author{X. Sheng}
\affiliation{Astrophysics Research Centre, School of Mathematics and Physics, Queens University Belfast, Belfast BT7 1NN, UK}

\author[0000-0003-4494-8277]{G. P. Smith}
\affiliation{School of Physics and Astronomy, University of Birmingham, Birmingham B15 2TT, UK}

\author{G. Terreran}
\affiliation{Las Cumbres Observatory, 6740 Cortona Drive, Suite 102, Goleta, CA 93117-5575, USA}
\affiliation{Department of Physics, University of California, Santa Barbara, CA 93106-9530, USA}

\author{M. A. Tucker}
\altaffiliation{CCAPP Fellow}
\affiliation{Department of Astronomy, The Ohio State University, 140 West 18th Avenue, Columbus, OH, USA}
\affiliation{Department of Physics, The Ohio State University, 191 West Woodruff Ave, Columbus, OH, USA}
\affiliation{Center for Cosmology and Astroparticle Physics, The Ohio State University, 191 West Woodruff Ave, Columbus, OH, USA}

\author{J. Vink\'o}
\affiliation{Konkoly Observatory, CSFK, MTA Centre of Excellence, Konkoly Thege M. \'ut 15-17, Budapest, 1121, Hungary}
\affiliation{ELTE E\"otv\"os Lor\'and University, Institute of Physics and Astronomy, P\'azm\'any P\'eter s\'et\'any 1/A, Budapest, 1117 Hungary}
\affiliation{Department of Experimental Physics, University of Szeged, D\'om t\'er 9, Szeged, 6720, Hungary}
\affiliation{Department of Astronomy, University of Texas at Ausin, 2515 Speedway Stop C1400, Austin, TX, 78712-1205, USA}

\author{R. J. Wainscoat}
\affiliation{Institute for Astronomy, University of Hawaii, 2680 Woodlawn Drive, Honolulu HI 96822, USA}

\author[0000-0002-3073-1512]{P. Wiseman}
\affiliation{School of Physics and Astronomy, University of Southampton, Southampton, SO17 1BJ, UK}

\author[0000-0002-1229-2499]{D. R. Young}
\affiliation{Astrophysics Research Centre, School of Mathematics and Physics, Queens University Belfast, Belfast BT7 1NN, UK}




\begin{abstract}
We present the discovery and extensive follow-up of a remarkable fast-evolving optical transient, AT2022aedm, detected by the Asteroid Terrestrial impact Last Alert Survey (ATLAS). AT2022aedm exhibited a rise time of $9\pm1$ days in the ATLAS $o$-band, reaching a luminous peak with $M_g\approx-22$\,mag. It faded by 2 magnitudes in $g$-band during the next 15 days. These timescales are consistent with other rapidly evolving transients, though the luminosity is extreme. Most surprisingly, the host galaxy is a massive elliptical with negligible current star formation. X-ray and radio observations rule out a relativistic AT2018cow-like explosion. A spectrum in the first few days after explosion showed short-lived He II emission resembling young core-collapse supernovae, but obvious broad supernova features never developed; later spectra showed only a fast-cooling continuum and narrow, blue-shifted absorption lines, possibly arising in a wind with $v\approx2700$\,\kms. We identify two further transients in the literature (Dougie in particular, as well as AT2020bot) that share similarities in their luminosities, timescales, colour evolution and largely featureless spectra, and propose that these may constitute a new class of transients: luminous fast-coolers (LFCs). All three events occurred in passive galaxies at offsets of $\sim4-10$\,kpc from the nucleus, posing a challenge for progenitor models involving massive stars or massive black holes. The light curves and spectra appear to be consistent with shock breakout emission, though usually this mechanism is associated with core-collapse supernovae. The encounter of a star with a stellar mass black hole may provide a promising alternative explanation.
\end{abstract}

\keywords{Transient sources (1851) --- Supernovae (1668) --- Tidal disruption (1696)}

\section{Introduction} \label{sec:intro}


Astrophysical transients are now found in their thousands by optical time-domain surveys with wide-field robotic telescopes, such as the Asteroid Terrestrial impact Last Alert System \citep[ATLAS;][]{Tonry2018}, Panoramic Survey Telescope and Rapid Response System \citep[Pan-STARRS;][]{Chambers2016}, Zwicky Transient Facility \citep[ZTF;][]{Bellm2019}, and All-sky Automated Search for Supernovae \citep[ASAS-SN;][]{Shappee2014}. These are unearthing a variety of new phenomena, and survey power is now set to increase even further with the Rubin Observatory, the first wide-field survey on an 8m-class telescope \citep{Ivezic2019}.


Improvements in survey cadence allow us to probe populations of transients that rise and fade on timescales of days, compared to the weeks-months of typical supernovae (SNe). 
The majority of fast transients seem to arise from stripped massive stars \citep{Ho2023}. These include the initial cooling peaks of Type IIb SNe, as well as events strongly interacting with a dense circumstellar medium (CSM) deficient in hydrogen \citep[SNe Ibn;][]{Pastorello2007,Hosseinzadeh2017} and sometimes helium \citep[SNe Icn;][]{Fraser2021,Gal-Yam2022,Perley2022}.

A more mysterious population of fast transients has also been uncovered, with blue colours and a wide range of peak luminosities up to $M<-20$\,mag, approaching superluminous supernovae \citep[SLSNe;][]{Quimby2011,Gal-Yam2019a}. Since their identification in Pan-STARRS by \citet{Drout2014}, such objects have been discovered in data from the Palomar Transient Factory and the Supernova Legacy Survey \citep{Arcavi2016}, the Dark Energy Survey \citep{Pursiainen2018}, ATLAS \citep{Prentice2018}, \textit{Kepler} \citep{Rest2018}, Hyper Suprime Cam \citep[HSC;][]{Tampo2020,Jiang2022} and ZTF \citep{Ho2023}. They have been termed Fast Blue Optical Transients (FBOTs) or Rapidly Evolving Transients (RETs). Their association with star-forming galaxies suggests a connection with massive stars \citep{Wiseman2020}, and their photometric evolution appears consistent with shock breakout from a dense, extended envelope \citep{Drout2014,Pursiainen2018}. 

Several of the best observed RETs also show persistent high temperatures and luminous X-ray and radio emission. The most famous example is AT2018cow \citep{Prentice2019,Margutti2019, Perley2019}, but other well-studied objects include CSS161010 \citep{Coppejans2020}, AT2018lug \citep{Ho2020}, AT2020xnd \citep{Perley2021,Ho2022,Bright2022}, and AT2020mrf \citep{Yao2022a}. These `Cow-like' events seem to be energised by continuous injection from a central engine \citep{Margutti2019}, though interaction with CSM can also contribute luminosity \citep{Ho2020}. Tidal disruption events (TDEs) of stars by intermediate-mass black holes (BHs) have been considered as an alternative model \citep{Perley2019}, though accretion onto a stellar-mass BH following the collapse of a massive star appears to be favoured by most authors \citep[e.g][]{Perley2021,Coppejans2020,Gottlieb2022}.


Here we present an extraordinary new rapid transient that points to a distinct class of luminous, fast-cooling events. AT2022aedm, or ATLAS22bonw, was discovered by ATLAS on 30 Dec 2022 \citep{2022aedm_disc}. Spectroscopy carried out the following day by the Advanced Public ESO Spectroscopic Survey of Transient Objects \citep[ePESSTO+;][]{Smartt2015} suggested a likely SN, though of indeterminate spectral type \citep{2022aedm_class}. A spectroscopic host galaxy redshift $z=0.14343$ from the Sloan Digital Sky Survey \citep[SDSS;][]{Albareti2017} indicated a peak absolute magnitude $M_o=-21.5$\,mag\footnote{We assume a flat $\Lambda$CDM cosmology with $H_0=70$\,\kms\,Mpc$^{-1}$ and $\Omega_\Lambda=0.7$, giving a luminosity distance $D_L=679$\,Mpc, and a Galactic extinction $E(B-V)=0.0428$ for this line-of-sight \citep{Schlafly2011}}, but the rising light curve was faster than any known SLSN. A relatively featureless spectrum, and most surprisingly an elliptical host galaxy, further added to the intrigue of this event, motivating extensive follow-up observations.

\begin{figure*}[ht!]
\centering
\includegraphics[width=8.9cm]{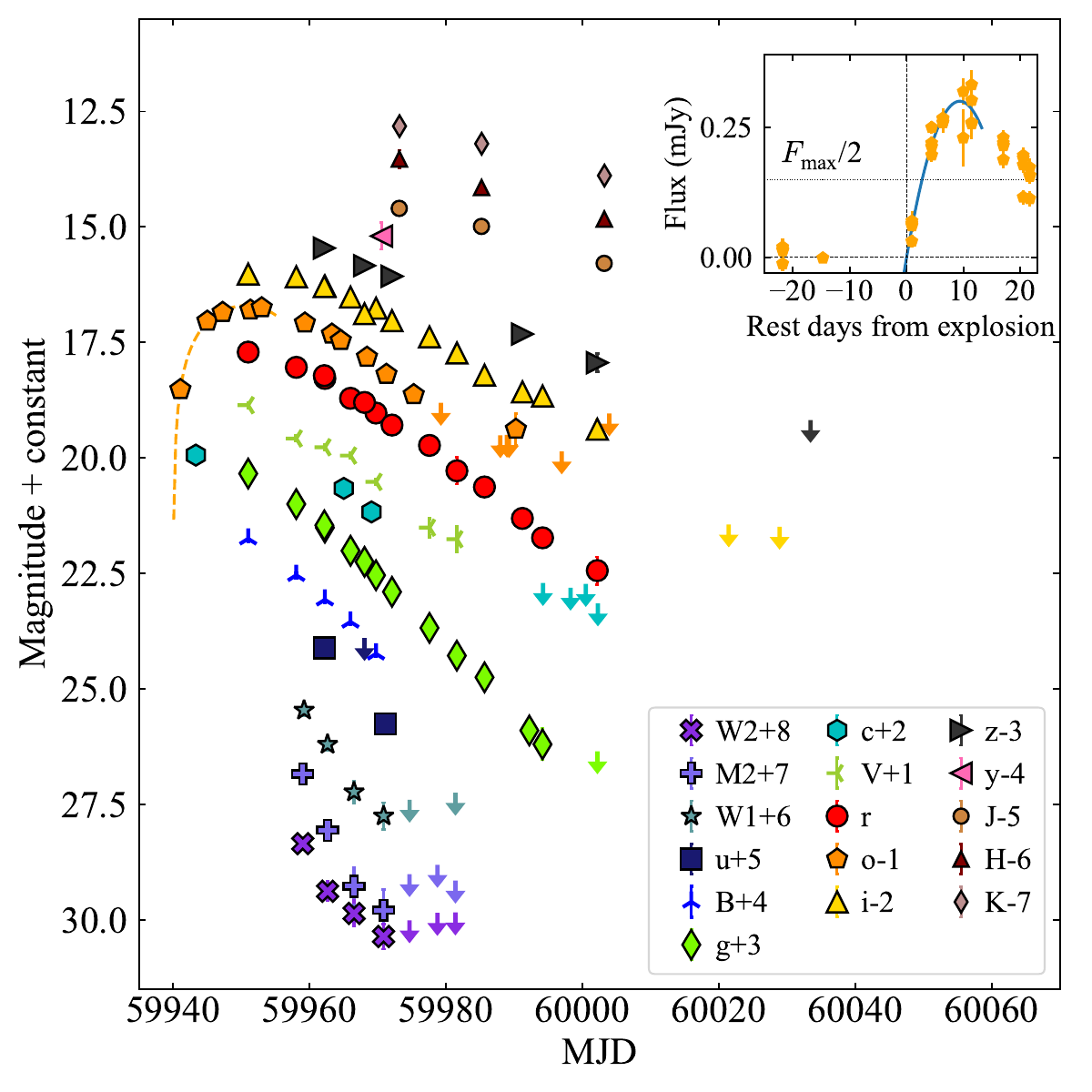}
\includegraphics[width=8.9cm]{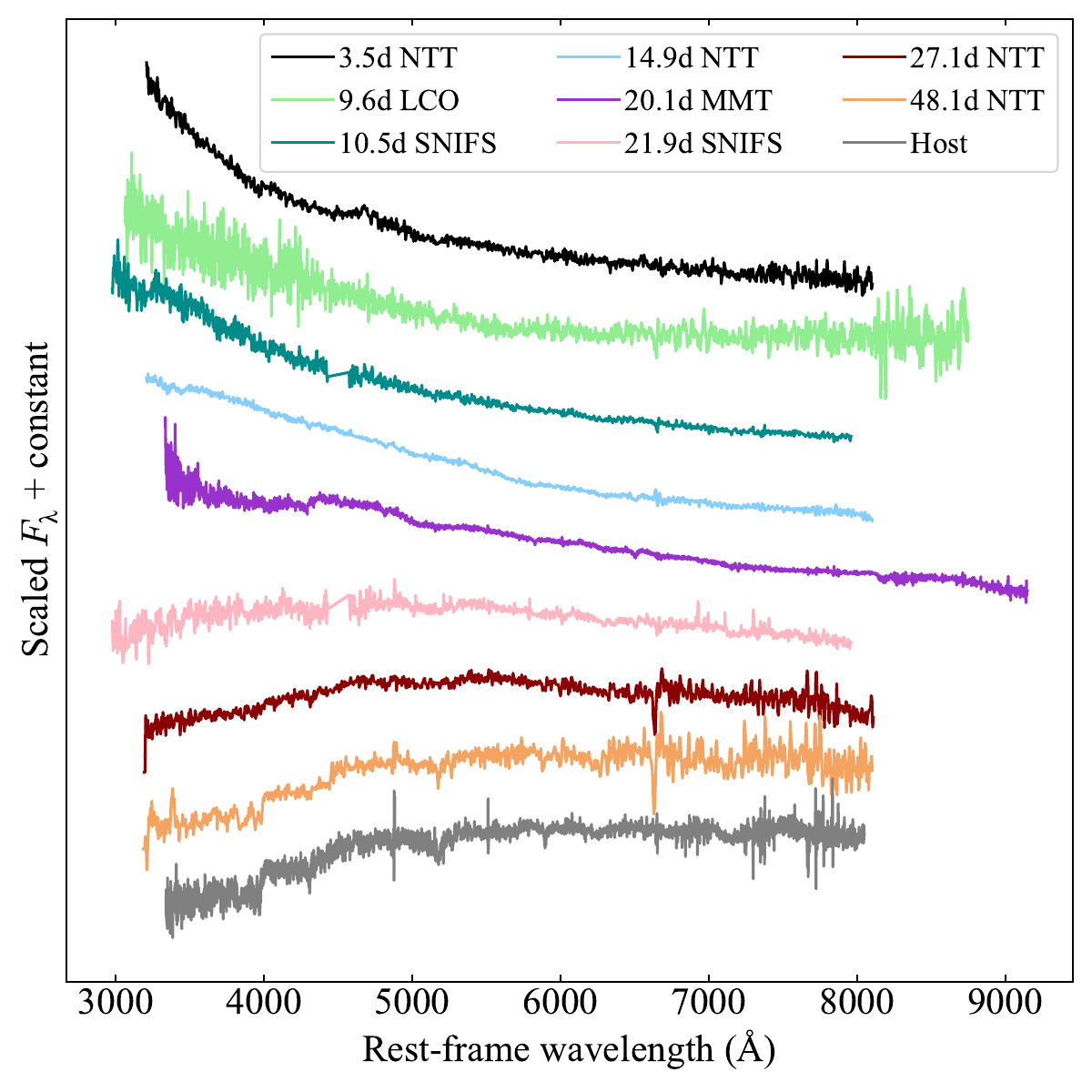}
\caption{
Follow-up of AT2022aedm. Left: multi-colour light curves from UV to NIR. Arrows indicate upper limits. The inset shows the early ATLAS flux data with a second-order fit (also shown on the multi-colour plot) and time above half-maximum. Taking the shallowest and steepest rises allowed by the ATLAS data changes the time of zero flux by $\lesssim 0.5$\,days. Right: spectroscopic follow-up. Spectra are labelled by telescope/instrument and time since explosion in the rest frame.
\label{fig:data}}
\end{figure*}

\section{Observations}

\subsection{Ground-based imaging}

AT2022aedm was discovered in the ATLAS transient stream processed by the ATLAS Transient Science Server \citep{KSmith2020}. 
Calibrated ATLAS data in the cyan ($c$) and orange ($o$) bands were obtained using the ATLAS forced photometry service \citep{Tonry2018,KSmith2020,Shingles2021}. ATLAS typically obtains four exposures per night in a given band; we combined each quad into a single average flux measurement to improve the signal-to-noise ratio. After AT2022aedm had faded below $o\sim19$\,mag, observations were binned over neighbouring nights to improve the signal-to-noise ratio.

Follow-up photometry was obtained from Pan-STARRS in the $i,z,y$ bands, the Las Cumbres observatory global telescope network (LCO, as part of the Global Supernova Project) in the $B,g,V,r,i$ bands, the Liverpool Telescope (LT) in the $u,g,r,i,z$ bands. Data from the European Southern Observatory New Technology Telescope (NTT) were obtained using both EFOSC2 for the optical $g,r,i,z$ bands and SOFI for the near-infrared (NIR) $J,H,K$ bands, as part of ePESSTO+. Data were reduced (de-biased and flat-fielded) either automatically by facility pipelines, or manually using the PESSTO pipeline \citep{Smartt2015} in the case of the NTT data.

\subsection{Photometry sans frustration}

Photometry was performed using a custom pipeline, `Photometry Sans Frustration' (or \textsc{psf})\footnote{\url{https://github.com/mnicholl/photometry-sans-frustration}.}. This is a fully \textsc{python}-based code, employing aperture and Point-spread function (PSF) fitting photometry routines from \textsc{astropy} \citep{Astropy2013,Astropy2018} and \textsc{photutils} \citep{Bradley2020}. As well as deriving the image background, PSF and zeropoint, the code provides options to automatically download local star catalogs and reference images using \textsc{astroquery} \citep{Ginsburg2019}, solve the coordinate system using \textsc{astrometry.net} \citep{Lang2010}, clean cosmic rays using \textsc{lacosmic} \citep{vanDokkum2012}, align and stack images using \textsc{astroalign} \citep{Beroiz2020}, and subtract transient-free reference images of the field using \textsc{pyzogy} \citep{Zackay2016,Guevel2017}.

All NTT, LCO and LT images were first cleaned and stacked within each night. Calibration stars and template images were obtained in $g,r,i,z$ from Pan-STARRS \citep{Flewelling2020}, in $u$ from the SDSS, and in $J,H,K$ from the VISTA Kilo-degree Galaxy Survey \citep[VIKINGS;][]{Edge2013}. LCO $B,V$ reference images were obtained after the transient faded. {The PSF for each science and template image was determined using the local reference stars. All transient fluxes were measured after template subtraction to remove host galaxy light.} The full photometric data set is shown in Figure \ref{fig:data}.

\subsection{Swift observations}

We obtained ultraviolet (UV) and X-ray data using target of opportunity observations with the UV-Optical Telescope (UVOT; \citealt{Roming2005}) and X-ray Telescope (XRT; \citealt{Burrows2005}) on-board the Neil Gehrels Swift Observatory (\textit{Swift}; \citealt{Gehrels2004}). UVOT imaging was carried out in the $uvw2$, $uvm2$ and $uvw1$ filters. The light curves were measured using a $5''$ aperture. Count rates were obtained using the \textit{Swift} \textsc{uvotsource} tools and converted to magnitudes (in the AB system) using the UVOT photometric zero points \citep{Breeveld2011}. No host subtraction was performed in the UV bands, as host contamination is negligible at these wavelengths (this is confirmed by the later UVOT visits that result in only non-detections).

We processed the XRT data using the online analysis tools provided by the UK \textit{Swift} Science Data Centre \citep{Evans2007,Evans2009}. AT2022aedm is not detected in the combined $15.3$\,ks exposure, with a limiting count rate $<8.53\times10^{-4}$\,s$^{-1}$. Assuming a power-law spectrum with $\Gamma=2$ \citep{Coppejans2020}, and a Galactic hydrogen column density towards AT2022aedm of $4.2\times10^{20}$\,cm$^{-2}$ \citep{HI4PI2016}, this corresponds to an unabsorbed 0.3-10\,keV luminosity $L_X<3.8\times10^{42}$\,\ergs. This upper limit is deeper than the observed X-ray luminosities $\sim10^{43-44}$\,\ergs\ in AT2018cow \citep{Margutti2019}, AT2020xnd \citep{Ho2022} and AT2022tsd \citep{Matthews2023}.

\subsection{Radio observations}

We observed AT2022aedm with the Arcminute Microkelvin Imager - Large Array \citep[AMI--LA;][]{Zwart2008, hickish2018} over 4 epochs, beginning 20 days after discovery. AMI--LA is an eight-dish interferometer based in Cambridge, UK. Each dish is 12.8\,m in diameter enabling an angular resolution of $\sim30"$. The facility observes at a central frequency of 15.5\,GHz with a bandwidth of 5\,GHz. 

Data taken with AMI--LA were reduced using a custom pipeline \textsc{reduce\_dc}. We used 3C286 and J1119$+$0410 as the primary and secondary calibrators, respectively, to perform amplitude and phase calibration. The pipeline also flags the data for radio-frequency interference, effects of poor weather and antenna shadowing. The data were then exported in \textit{uvfits} format ready for imaging. 

Further flagging and imaging were conducted in the Common Astronomy Software Applications (\textsc{casa}) \citep[Version 4.7.0][]{McMullin2007, casa} using the tasks \textit{tfcrop}, \textit{rflag} and \textit{clean}. AT2022aedm is not detected in any of the final images, with 3$\sigma$ upper limits of $F_{\nu}<[41,210,156,96]\,\mu$Jy at 20, 66, 107 and 110 days after the first optical detection. These correspond to limits on the spectral luminosity of $L_{\nu}<2.3\times10^{28}-1.2\times10^{29}$\,\ergs\,Hz$^{-1}$. For comparison, Cow-like RETs typically exhibit radio emission at the level of $L_{\nu}\gtrsim10^{29}$\,\ergs\,Hz$^{-1}$ (at $\sim10$\,GHz) on timescales of months \citep{Coppejans2020,Ho2020}.

\begin{figure*}[ht!]
\centering
\includegraphics[width=8.9cm]{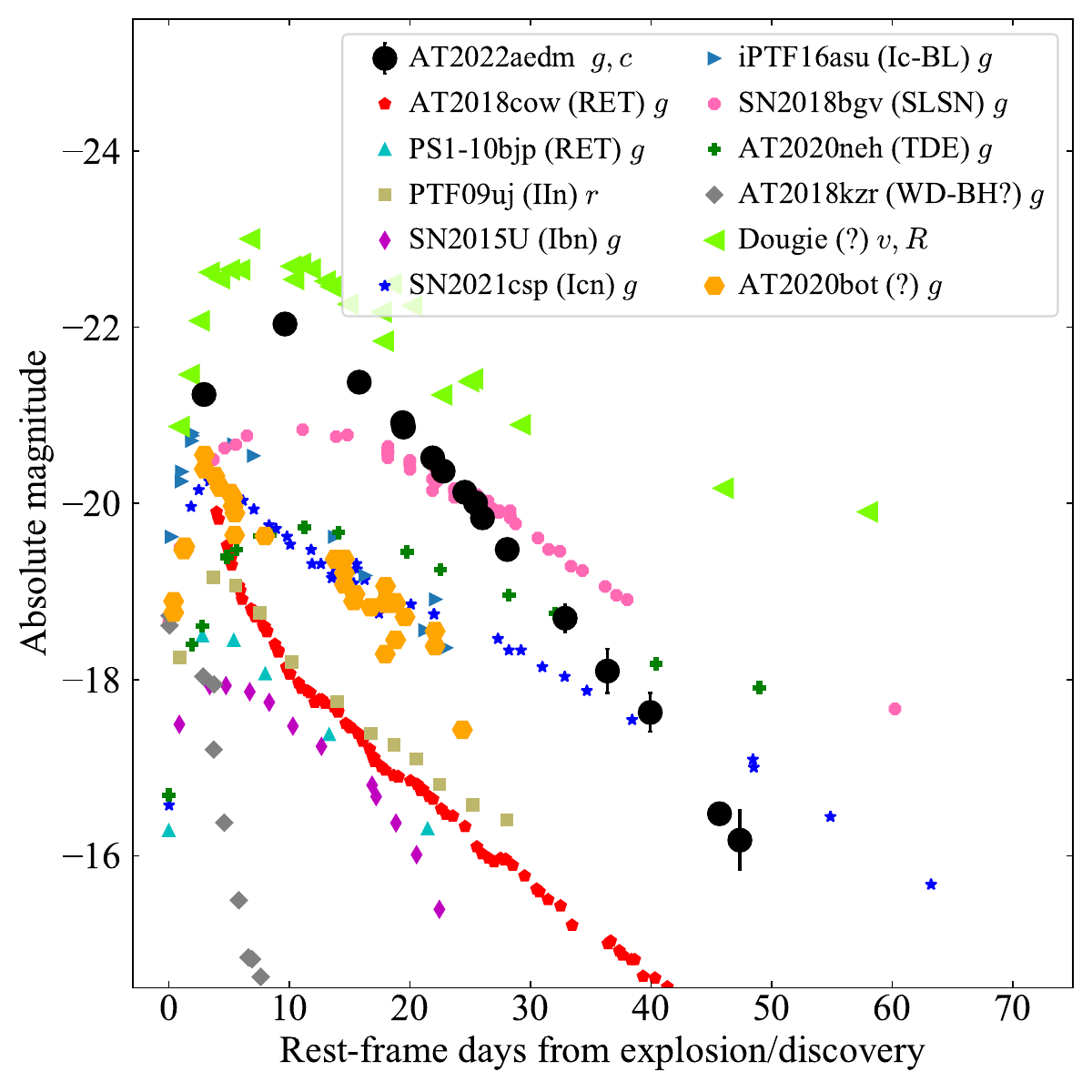}
\includegraphics[width=8.9cm]{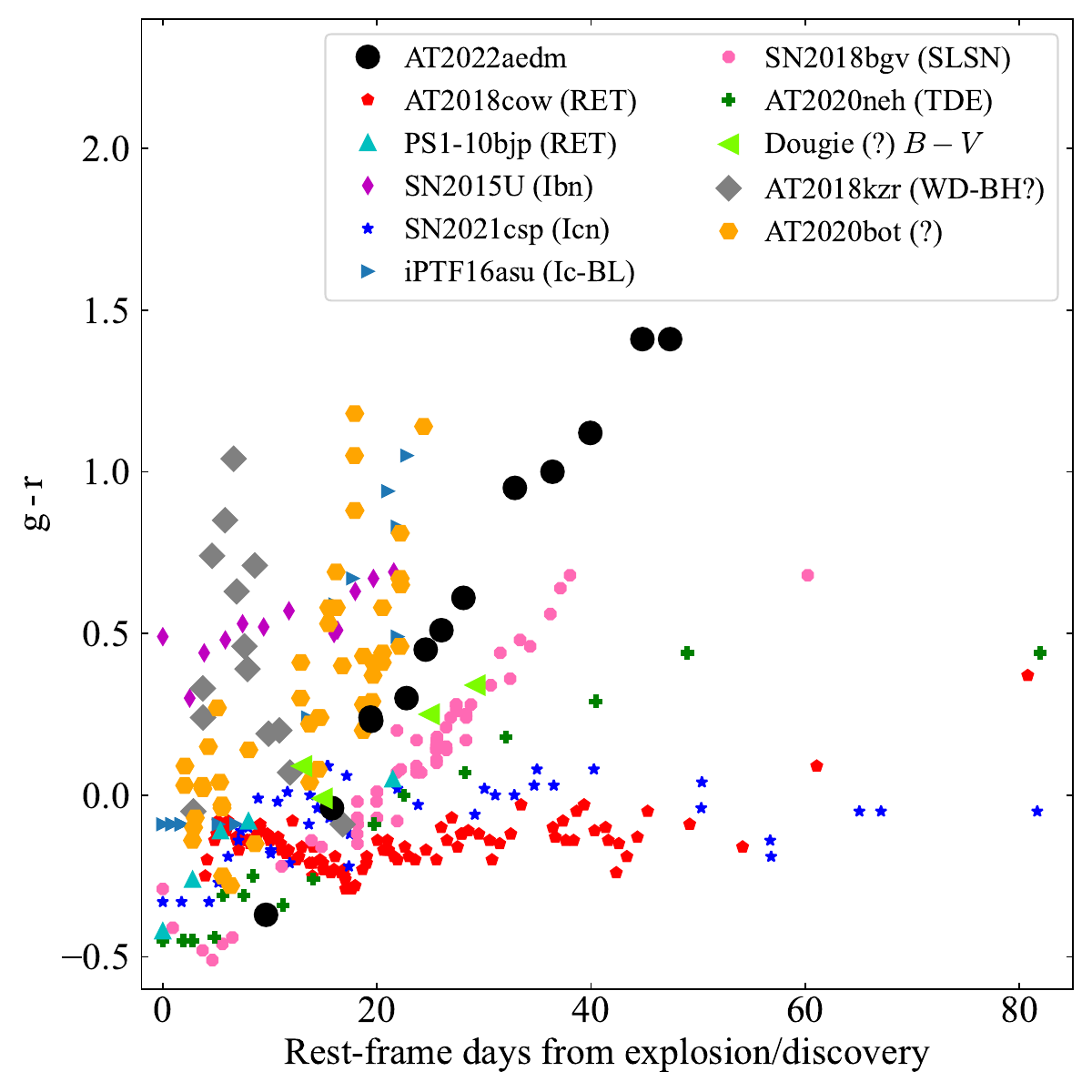}
\caption{
Photometric comparison of AT2022aedm and other fast transients. For visual clarity, we plot the uncertainties only for AT2022aedm. Left: absolute rest-frame light curves. Right: $g-r$ colour evolution.
\label{fig:lc}}
\end{figure*}

\subsection{Spectroscopy}

We obtained optical spectra of AT2022aedm using EFOSC2 on the NTT (through ePESSTO+), the LCO 2-m telescopes, the SuperNova Integral Field Spectrograph \citep[SNIFS;][]{Lantz2004,Tucker2022} on the University of Hawaii 2.2-m telescope, and Binospec on the 6.5-m MMT \citep{Fabricant2019}. Spectroscopy from ePESSTO+ commenced on 31 Dec 2022 (within one day after the object was flagged by ATLAS) and continued until 20 Feb 2023, by which time the spectrum was indistinguishable from a pre-explosion host galaxy spectrum from SDSS. 

Standard reductions of these data, including de-biasing, flat-fielding, spectral extraction, flux and wavelength calibration, were performed using instrument-specific pipelines. The reduced spectra are plotted in Figure \ref{fig:data} and labelled with the instrument and phase with respect to our estimated explosion date. We assume that host galaxy extinction is negligible, supported by the lack of Na\,ID absorption in these spectra \citep{Poznanski2012}, and the early blue colours in our spectra and photometry. All data will be made publicly available via WISeREP \citep{Yaron2012}.

\section{Analysis}

\subsection{Light curve}
\label{sec:phot}


The rising light curve of AT2022aedm is well constrained by the early ATLAS $o$-band detections. The discovery point on MJD 59941.1 at $o=19.51$\,mag is a factor $\simeq5$ below the peak $o$-band flux. Fitting a second-order polynomial to the early flux light curve (Figure \ref{fig:data}) indicates the explosion occurred on MJD $59940.0\pm0.5$ ($\approx1$ day before the first detection), reaching $o$-band peak on MJD $59950.6\pm0.5$. We take these as the dates of explosion and peak throughout. We note however that the last non-detection is 15 days before detection, so a slightly earlier explosion date cannot be entirely ruled out if the early light curve shape is more complex. The rest-frame rise-time from half the peak flux is $t_{r,1/2}=6.6$\,days, much shorter than the SLSNe that reach comparable peak luminosities but with $t_{r,1/2}=10-40$ days \citep{Nicholl2015,DeCia2018,Lunnan2018a}.


The fading timescale of AT2022aedm is also much quicker than most other luminous transients. The $g$-band light curve fades by 2 magnitudes in the 15 days after peak, and by 18 days has declined to 10\% of peak $g$-band flux. In $o$-band, where we also have the rise, the full-width at half-maximum (i.e.~the total time spent within 50\% of peak flux) is $t_{1/2}=19\pm1$ days.
These timescales are well within the distributions measured for RETs discovered in DES \citep{Pursiainen2018}. Although the measured $t_{1/2}$ for AT2022aedm is longer than the $t_{1/2}<12$\,days defining RETS in PS1 \citep{Drout2014} and ZTF \citep{Ho2023}, this is attributable to using different photometric filters: we measured $t_{1/2}$ in $o$, where the rate of fading in AT2022aedm is $\approx55\%$ slower than in the $g$-band. Correcting by this factor gives an estimated $g$-band $t_{1/2}$ of $\approx12$ days.

\begin{figure*}[ht!]
\centering
\includegraphics[width=8.9cm]{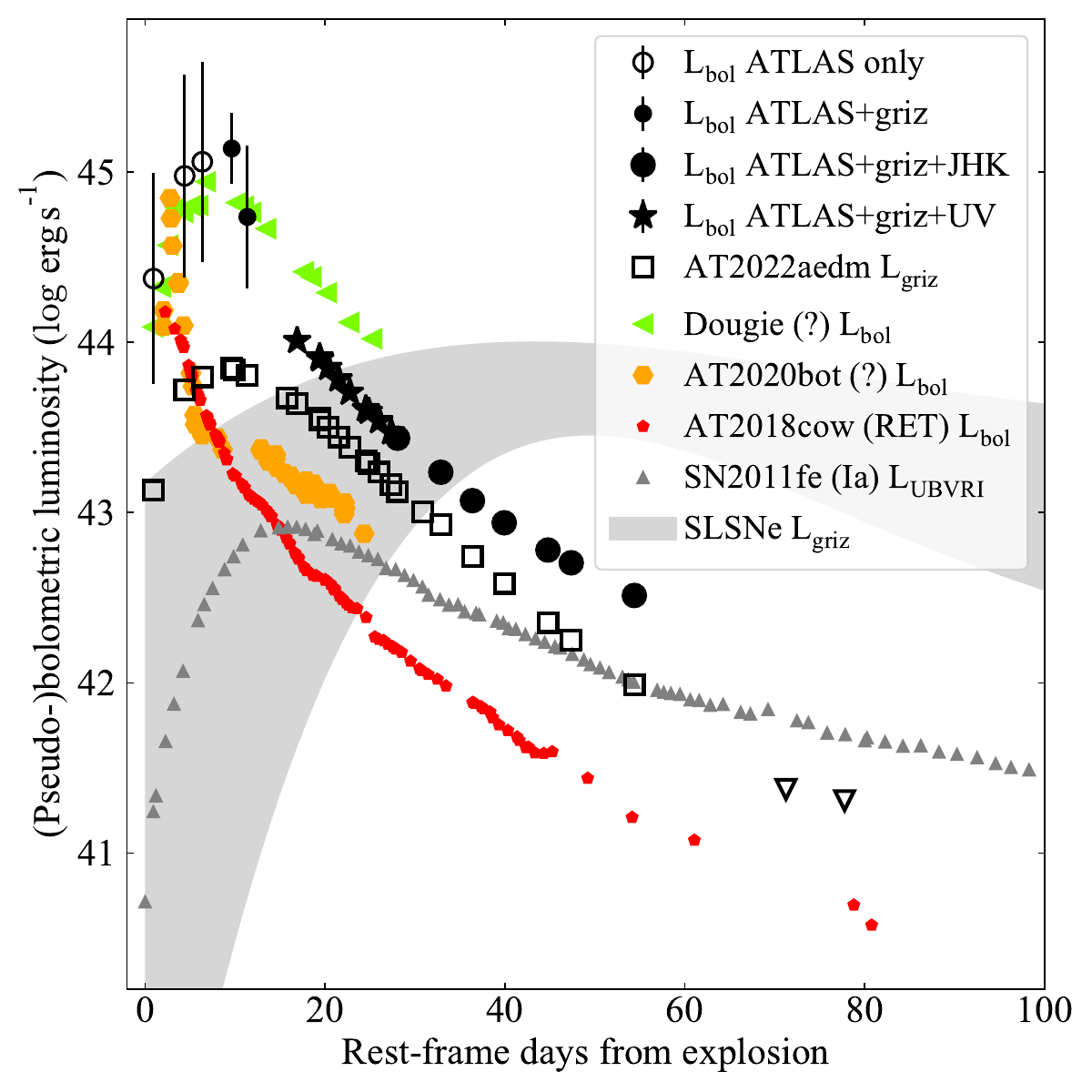}
\includegraphics[width=8.9cm]{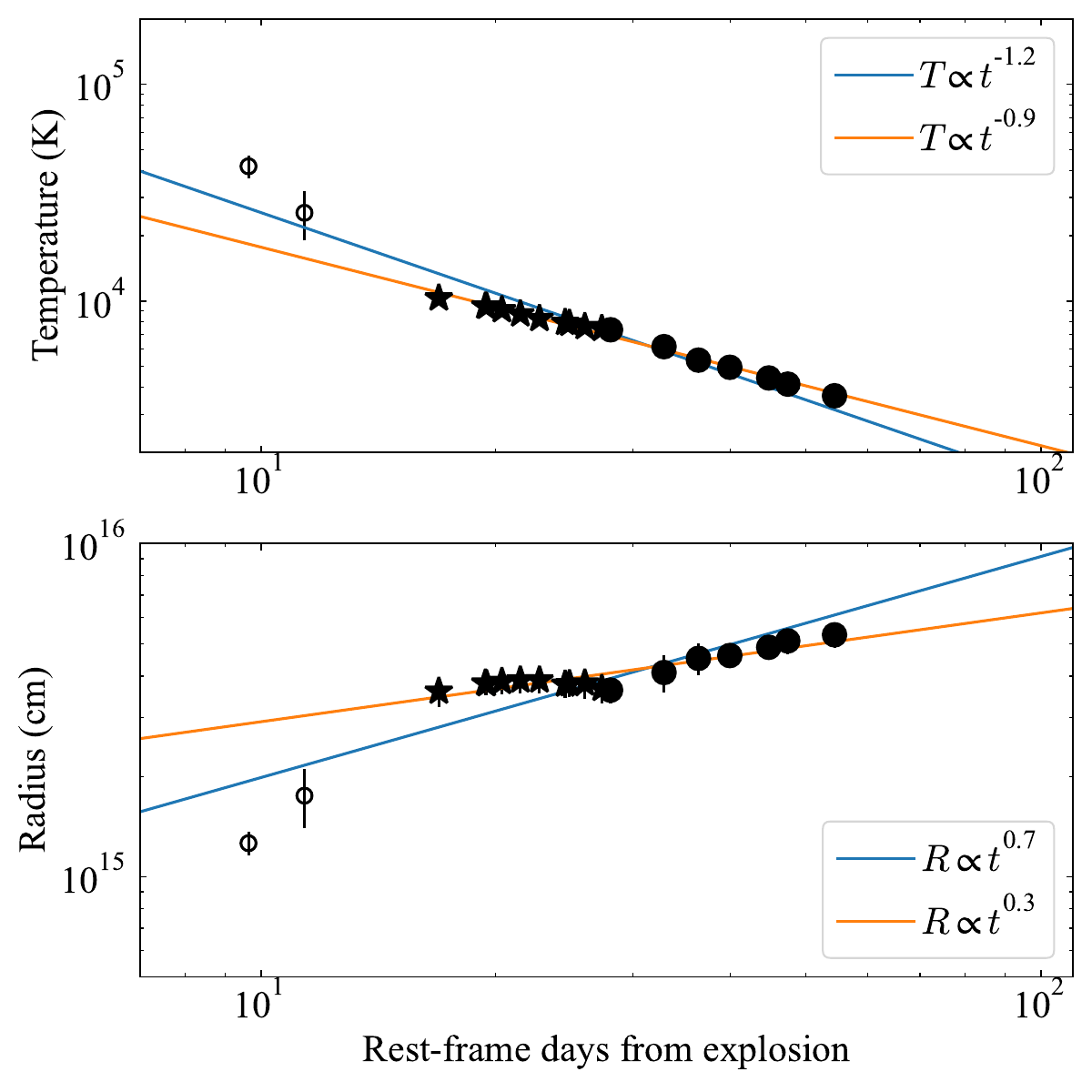}
\caption{
Results from fitting the multi-band photometry with \textsc{superbol}. Left: bolometric and pseudo-bolometric (optical) light curves compared to AT2018cow, AT2020bot and Dougie. We also show a representative SN Ia \citep{Nugent2011} and a SLSN sample \citep{Nicholl2015}. For visual clarity, we plot the uncertainties only for AT2022aedm. Right: Temperature and radius evolution from blackbody fits, together with best-fitting power-laws. The orange lines are not fit to the first two data points, for which no UV or NIR data are available.
\label{fig:bol}}
\end{figure*}


The extinction-corrected $g$-band peak luminosity of AT2022aedm, $M_g=-22.04\pm0.05$, makes it one of the brightest RETs discovered to date. It outshines all but one event (DES16E1bir) in the combined PS1+DES+ZTF sample. The closest spectroscopically-classified RETs in terms of luminosity are the Cow-like RETs, typically reaching $\approx -21$\,mag \citep{Ho2023}. To highlight the exceptional luminosity of AT2022aedm, we show a combined $g$- and $c$-band rest-frame light curve in Figure \ref{fig:lc}, compared to representative examples of different types of fast-fading transients. AT2022aedm is broader and brighter than AT2018cow, but fades faster than the fastest SLSN, SN2018bgv \citep{Chen2023a}. It is much more luminous than a typical RET \citep{Drout2014}, the fastest TDE, AT2020neh \citep{Angus2022}, the fastest broad-lined SNe Ic, such as iPTF16asu \citep{Whitesides2017} (see also SN2018gep and SN2018fcg; \citealt{Pritchard2021}; \citealt{Gomez2022}), or any fast interacting transients of Types IIn \citep{Ofek2010}, Ibn \citep{Hosseinzadeh2017} or Icn \citep{Fraser2021,Perley2022}.


Figure \ref{fig:lc} also shows the $g-r$ (or $B-V$) colour evolution of AT2022aedm compared to the same sample of objects (where multiple bands are available). From an initial $g-r=-0.37$ at 10 days after explosion, AT2022aedm dramatically reddens by 1.8 magnitudes in colour index over the next 35 rest-frame days. Cow-like RETs, TDEs, and interacting transients generally show a more gradual or flat colour evolution. The colour change in AT2022aedm is more consistent with events with expanding, cooling photospheres; it lies intermediate between iPTF16asu and SN2018bgv. PS1-10bjp shows a similar colour evolution over the first 10 days. 

Two unclassified fast transients show a comparable colour evolution, in combination with a peak absolute magnitude brighter than $-20$\,mag. One is AT2020bot, the only RET in the ZTF sample that did not fit into any of the stripped-envelope, interacting or Cow-like sub-populations \citep{Ho2023}. It is fainter than AT2022aedm, with a faster rise and redder average colour. A stronger similarity is exhibited by `Dougie' \citep{Vinko2015}, a mysterious transient discovered by ROTSE in 2009. This event peaked at $-22.5$\,mag after a fast rise of $\approx10$ days. The early light curve shape is very similar to AT2022aedm, though the decline may flatten after 30-40 days. However, this flattening could also be due to a host contribution in its UVOT photometry. While the light curve could be plausibly interpreted as a super-Eddington TDE \citep{Vinko2015}, its position offset from the host nucleus, and lack of distinct spectroscopic features, make this classification far from certain.

\subsection{Bolometric light curve}

To measure the overall energetics of AT2022aedm, we integrate our multi-band photometry using \textsc{superbol} \citep{Nicholl2018}. We construct a pseudo-bolometric light curve using the excellent $g,r,i,z$ and $o$-band coverage, and estimate the full bolometric light curve by fitting blackbody functions to these bands and to our UV and NIR photometry where available. These light curves are shown in Figure \ref{fig:bol}.

\begin{figure*}[ht!]
\centering
\includegraphics[width=8.9cm]{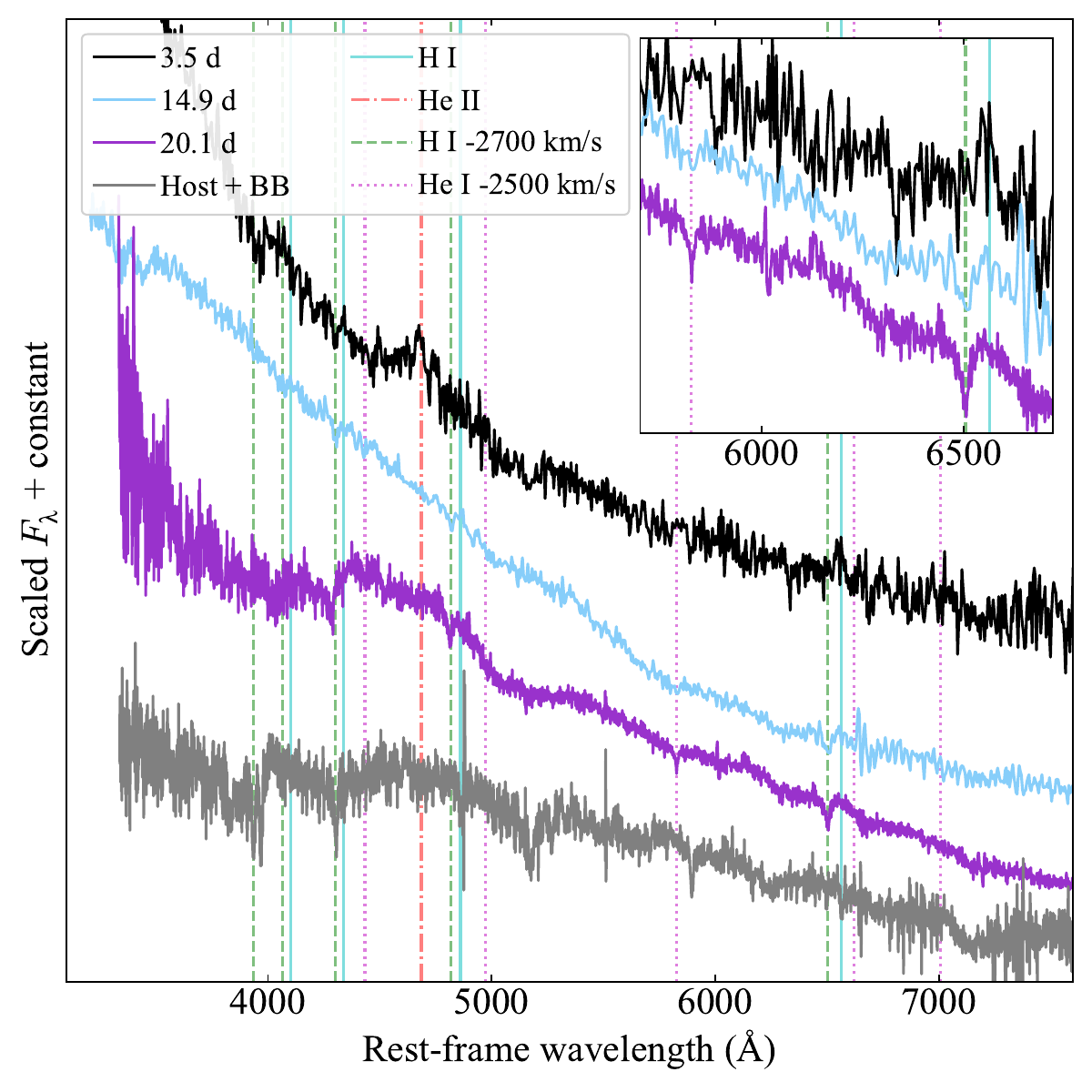}
\includegraphics[width=8.9cm]{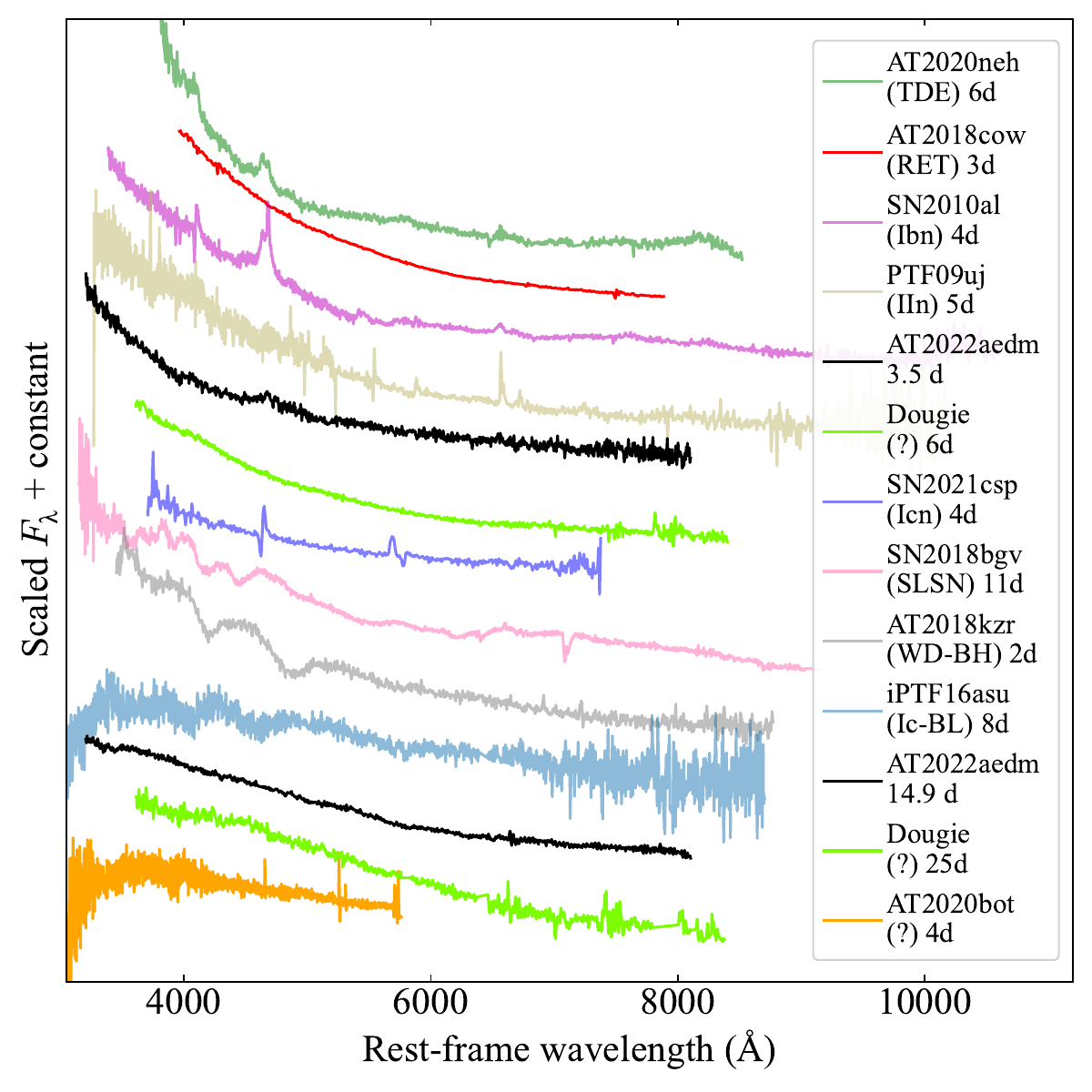}
\caption{
Spectroscopic analysis of AT2022aedm. Left: line identification in the highest signal-to-noise ratio spectra. No unambiguous broad features are identified, but we clearly detect early emission lines of H$\alpha$ and He~II, and at later times blueshifted absorption lines of H and He~I.
Right: comparison of AT2022aedm with other fast transients. Dougie and AT2020bot also show a lack of obvious SN features and occurred in similar environments to AT2022aedm.
\label{fig:spec}}
\end{figure*}

The peak pseudo-bolometric luminosity is typical of SLSNe, reaching $L_{griz}=10^{43.8}$\,\ergs, but the light curve rise and decline rates fall well outside the SLSN distribution. The decay rate is comparable to AT2018cow, though the rise is longer than the $<2$\,days exhibited by that event. The estimated full bolometric luminosity of AT2022aedm at peak is exceptionally high, reaching $\approx10^{45}$\,\ergs. This is due to a high temperature $T\gtrsim30,000$\,K (shown in the top right panel), suggested by the very blue $g-r$ and $r-i$ colours in the first LCO images. The temperature exhibits a monotonic decline with a scaling of roughly $T\propto 1/t$. The blackbody radius increases throughout our observations, suggestive of an expanding photosphere that remains optically thick. The estimated bolometric light curves of Dougie and AT2020bot, also constructed using \textsc{superbol}, peak at similar luminosities to AT2022aedm.

\subsection{Spectra}


The spectroscopic evolution of AT2022aedm is shown in Figure \ref{fig:data}. The cooling observed in the photometry is also evident in its spectra. The first NTT spectrum 3.5 days after explosion shows a strong blue continuum, which weakens and disappears by day 27. By day 48, the spectrum is indistinguishable from an archival SDSS spectrum of the host galaxy. 


The spectra mostly lack obvious broad emission, absorption or P Cygni lines typically seen in SNe. The spectra on days 3 and 14 (both from NTT) and 20 (from MMT) with the best signal-to-noise ratios are examined in more detail in Figure \ref{fig:spec}. Weak broad features may exist at around 4000-5000\,\AA\ after day 20, though these could also be caused by contamination from the host galaxy, which is around 2 magnitudes brighter than AT2022aedm at this phase. We show this explicitly by adding an arbitrary $20,000$\,K blackbody to the host, finding that this reasonably reproduces the overall shape of the day 20 spectrum.


Figure \ref{fig:spec} does show several \emph{narrow} spectral lines of a clearly transient nature. The day 3 spectrum shows a sharply peaked emission line consistent with He\,II~$\lambda4686$, possibly with a broader base. This line is often observed in very young SNe \citep{Gal-Yam2014,Khazov2016,Bruch2021} and in TDEs \citep{Gezari2012,Arcavi2014}, due to the high radiation temperatures capable of ionising helium. We also detect a weak narrow H$\alpha$ emission line. Other Balmer lines are not visible at this signal-to-noise. He\,II is not detected in any later spectra, likely because the temperature has fallen too low to maintain helium ionisation. 


\begin{figure*}[ht!]
\centering
\includegraphics[width=12cm]{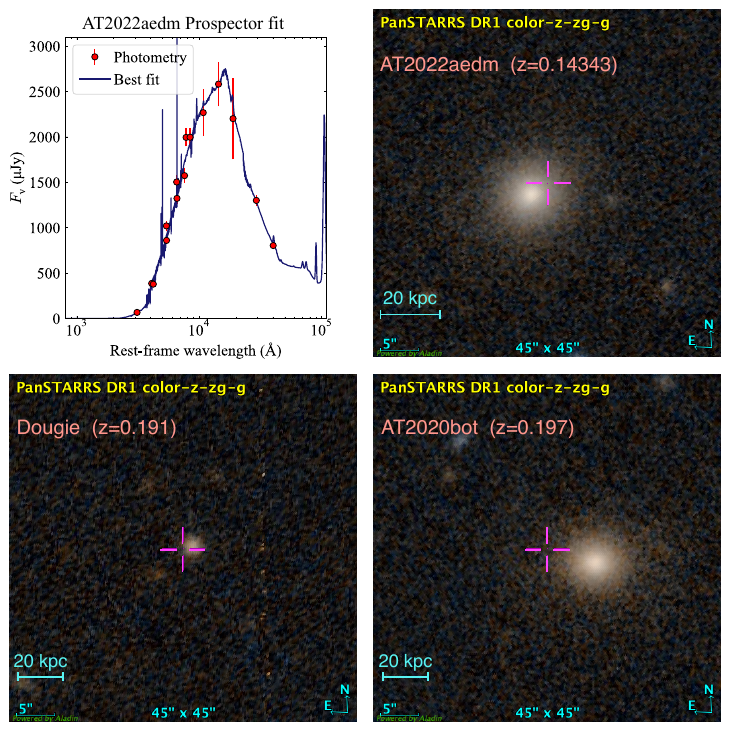}
\caption{
Top left: Fit to the AT2022aedm host galaxy SED using \textsc{Prospector}. Top right: Host galaxy colour image. Bottom row: The elliptical host galaxies and offsets of two other luminous fast-cooling transients with unusual spectra \citep{Vinko2015,Ho2023}. Images are from Pan-STARRS and centered at the transient positions.
\label{fig:host}}
\end{figure*}


The spectra on days 14 and 20 show narrow absorption, rather than emission, from hydrogen and neutral helium. The first three transitions of the Balmer series are clearly visible, but blue-shifted from their rest wavelengths by $\approx2700$\,\kms. The high-resolution MMT/Binospec data on day 20 also clearly show He\,I~$\lambda5875$ absorption, blue-shifted by $\approx2500$\,\kms. These lines are not visible in the host spectrum, confirming their association with the transient.


Figure \ref{fig:spec} also includes a comparison between the early spectra of AT2022aedm and other fast transients. The initial He\,II and H$\alpha$ emission is reminiscent of some SNe Ibn \citep{Hosseinzadeh2017}, as well as the fast-evolving TDE AT2020neh \citep{Angus2022}. The earliest spectrum is also a reasonable match for young SNe IIn, including the shock-breakout candidate PT09uj \citep{Ofek2010}. However, AT2022aedm never develops the strong emission lines typically seen in these classes at later times. The largely featureless spectrum even at 15 days is a poor match for any fast evolving SN Ic, SLSN, or SN Icn.

Cow-like RETs exhibit quite featureless spectra at peak, but the prototype AT2018cow showed increasingly clear H and He emission lines as it evolved, opposite to the case in AT2022aedm. Interestingly, the spectra of Dougie remained featureless for $\gtrsim30$\,days \citep{Vinko2015}, and seemed to cool in a manner similar to AT2022aedm. AT2020bot lacks a full spectroscopic time series, making a detailed comparison difficult. It shows an unusual spectrum at maximum light. \citet{Ho2023} note the presence of possible broad features, weak compared to typical SN lines and without an obvious identification. In section \ref{sec:phot} we observed that these two events also shared some key photometric properties with AT2022aedm.

\subsection{Host galaxy}


The host galaxy of AT2022aedm is LEDA 1245338 (or SDSS J111927.73+030632.7). This is a bright, red galaxy with $M_r=-22.8$\,mag; Figure \ref{fig:host} illustrates this with a colour image obtained from Pan-STARRS. An SDSS spectrum is also available (shown in Figure \ref{fig:spec}). Both the SDSS spectral fitting and Galaxy Zoo morphological analysis classify LEDA 1245338 as an elliptical galaxy.

The SDSS data release also includes automated analysis of the SDSS spectrum with the Portsmouth pipeline (using the method of \citealt{Maraston2009}). The spectral fitting measures a total stellar mass of $\approx10^{11.5}$\,\M, and a star-formation rate (SFR) consistent with zero. They find a mean age of the stellar population of 4.8\,Gyr. Given that SED modelling is highly sensitive to the assumed functional form of the star-formation history \citep{Carnall2019,Leja2019}, we run our own analysis on the host photometry, over a wider wavelength range, using \textsc{prospector} \citep{Johnson2021}. In particular, we use the \textsc{prospector}-$\alpha$ model employing a non-parametric star-formation history, with 6 equal-mass star-forming bins of flexible width \citep[see][for details]{Leja2017}. We include archival host photometry from SDSS, the 2 Micron All-Sky Survey \citep[2MASS;][]{Skrutskie2006}, and the Wide-field Infrared Survey Explorer \citep[WISE;][]{Wright2010}. The fit is shown in Figure \ref{fig:host}. We measure $M_*=10^{11.45}$\,\M, consistent with SDSS results, and a specific SFR in the last 50\,Myr of $\log ({\rm sSFR/yr}^{-1}) = -11.69$.

The SDSS analyses and our \textsc{prospector} results confirm that the host of AT2022aedm is a massive, `red and dead' galaxy. This is surprising: recent work by \citet{Irani2022a} shows that less than 1\% of core-collapse explosions occur in such environments. Moreover, this galaxy is especially unlike the hosts of most transients with comparable luminosity. SLSNe occur almost exclusively in low-mass galaxies with $< 10^{10}$\,\M\ \citep{Lunnan2014,Leloudas2015,Perley2016,Chen2017a,Schulze2018}, and their average specific star-formation rate is three orders of magnitude greater than our measurement for AT2022aedm. Bright RETs are also found in relatively low-mass, star-forming host galaxies: the sSFRs of the RET hosts in PS1 and DES span $-10\lesssim\log({\rm sSFR/yr}^{-1})\lesssim-8$, with only three (out of 73) fitting best to a passive galaxy model, and only two having a stellar mass greater than $10^{11}$\,\M. \citet{Wiseman2020} conducted a systematic analysis of RET hosts in DES, finding evidence for star-formation in all of the 49 galaxies for which redshifts were available, and the five RETs from HSC were also found in star-forming galaxies \citep{Tampo2020}.

Notably, two other fast transients from our photometric and spectroscopic comparisons also occurred in elliptical galaxies: Dougie and AT2020bot \citep{Vinko2015,Ho2023}. The Pan-STARRS images of their hosts are shown alongside AT2022aedm in Figure \ref{fig:host}. We also note that one SN Ibn, PS1-12sk, exploded in a bright elliptical, prompting \citet{Hosseinzadeh2019} to suggest that not all SNe Ibn result from massive stars.

\section{Discussion}

\subsection{A new class of transient?}

AT2022aedm is a puzzling event with a very unusual set of properties:
\begin{itemize}
    \item a high peak luminosity in the optical, with $M_g\approx-22$\,mag
    \item no luminous radio or X-ray emission
    \item fast rise and decline rates, fading $\sim 1$\,mag per week in the $g$-band
    \item rapid cooling from $\sim30,000$\,K to $\sim4,000$\,K in a few weeks following peak
    \item a spectrum dominated by a smooth continuum with no high equivalent width absorption or emission lines at any phase
    \item a massive host galaxy comprised of an old stellar population, with no evidence for current star formation.
\end{itemize}




This combination is not consistent with any known class of transients. 
An extensive search of the literature reveals two other objects which share the key properties: $M<-20$\,mag, rise time $<10$ days, fast decline and colour evolution, and weak spectroscopic features at all times. Together, these events indicate a new class of fast transients with high optical luminosities and fast cooling after peak. Dougie \citep{Vinko2015} in particular shows a strong photometric and spectroscopic similarity, while AT2022bot \citep{Ho2023} may represent an even faster-evolving member of this class. All three occurred off-centre within passive host galaxies, indicating that -- uniquely among RETs -- these new `Luminous Fast Coolers' (LFCs) do not require young stellar populations.

\subsection{Rates}


We estimate the volumetric rate of these events using the ATLAS Survey Simulator \citep{McBrien2021b,Srivastav2022}. Interpolated absolute light curves of AT2022aedm in $c$ and $o$ bands are inserted into the simulation at 10,000 random times, positions and redshifts (up to some $z_{\rm max}$), and compared to the cadence, footprint and depth of the true ATLAS survey to determine the number of $5\sigma$ detections of each injected event. 

We define a discovery as those objects detected in more than $n_{\rm det}$ images (or $\approx n_{\rm det}/4$ nights, since ATLAS typically obtains a quad of exposures per night at a given pointing). We then calculate the fraction classified of all real transients brighter in apparent magnitude than the injected transient at $z=z_{\rm max}$ and having at least the same number of ATLAS detections, and take this as our spectroscopic completeness. The rate is then 
\begin{equation}
    R= \frac{N_{\rm events}}{f_{\rm disc}f_{\rm spec} \int_{0}^{z_{\rm max}} T/(1+z)\,\frac{dV}{dz} dz}, 
\end{equation}
where $T=2.5$\,yr is the duration of the mock survey {(with the factor $1/(1+z)$ accounting for time dilation over the observed redshift range), $dV/dz$ is the differential comoving volume between redshift $z$ and $z+dz$}, $f_{\rm disc}$ and $f_{\rm spec}$ are the fractions discovered and classified, and $N_{\rm events}=1$ is the number of observed AT2022aedm-like transients in ATLAS.

We set $z_{\rm max}=0.2$, covering the redshift range within which LFCs have been discovered, and set $n_{\rm det}=20$. The latter is motivated by the modal number of detections for real ATLAS transients, and with observations on 5 nights should also enable identification of the fast light curve shape. Varying $n_{\rm det}$ leads to less than a factor two variation in our derived rate, due to a trade-off between $f_{\rm disc}$ and $f_{\rm spec}$: stricter requirements lead to a smaller discovered sample but with a higher spectroscopic completeness. For these parameters, our survey simulation returns $f_{\rm disc}=0.35$ and $f_{\rm spec}=0.58$, giving $R\approx1$\,Gpc$^{-3}$\,yr$^{-1}$.

We caution that this estimate applies to the brightest LFCs such as AT2022aedm and Dougie, and that fainter and faster events such as AT2020bot are likely more common volumetrically but harder to detect. Nevertheless, our derived rate indicates that these events are very rare, $\sim10^{-5}$ of the core-collapse SN rate. This rate is lower than the SLSN rate of a few\,$\times10$\,Gpc$^{-3}$\,yr$^{-1}$ \citep{Frohmaier2021}, but may be consistent with the rate of Cow-like events, estimated as $0.3-420$\,Gpc$^{-3}$\,yr$^{-1}$ \citep{Ho2023}.

\subsection{Physical scenarios for LFCs}

\subsubsection{Tidal disruption events}


\citet{Vinko2015} favoured a TDE as the origin of Dougie. A model with a relatively low-mass BH of $\sim10^5$\,\M\ provided a good match to the fast evolving light curve. Although the host galaxy luminosity was more consistent with a central BH mass of $\sim10^7$\,\M, Dougie's location $\sim 4$\,kpc from the nucleus could indicate a disruption around a wandering intermediate-mass BH.

This scenario has difficulties accounting for AT2022aedm. We are unable to find an acceptable fit using the TDE model in \textsc{mosfit}\footnote{This is an updated version of the same model used to fit Dougie; see \citet{Guillochon2014}.} \citep{Guillochon2018a,Mockler2019}, where models cannot reproduce the steep decline from peak or the fast colour evolution. The shallower decay of Dougie at late times may include some host contribution, causing a flattening that mimics the power-law decay of TDE models.

Other circumstantial problems arise in trying to explain LFCs as TDEs. AT2022aedm and AT2020bot have even larger offsets from the nuclei of their hosts. In particular, AT2020bot shows an offset of $\gtrsim10$\,kpc, in the outskirts of the galaxy where the stellar density is low. This would make a TDE very unlikely (though at available imaging depths we cannot rule out a globular cluster). TDE models would also need to explain why these offset events show such a strong evolution in colour compared to TDEs in their host nuclei, and are so much brighter than other TDEs with fast evolution \citep{Blagorodnova2017,Nicholl2020,Charalampopoulos2022,Angus2022}.

\subsubsection{Nickel powering and white dwarf explosions}


Most SNe are heated by the decay of radioactive nickel ($^{56}$Ni) to cobalt and then iron, but this mechanism can be excluded for many of the known RETs \citep[e.g.][]{Drout2014,Pursiainen2018,Perley2021}. The problem is that fast light curves require low ejecta masses ($M_{\rm ej}$), while bright peak luminosities require large nickel masses ($M_{\rm Ni}$). \citet{Drout2014} showed that to produce a peak luminosity of $\sim10^{44}$\,erg and a rise time $<10$\,days, a nickel-powered model would need an ejecta velocity $>0.1c$. We can probably rule out a very relativistic explosion in the case of AT2022aedm, due to our radio non-detections, though cannot exclude a mildly relativistic expansion. More problematic is the requirement for $M_{\rm ej}\sim M_{\rm Ni}\sim1$\M. This would produce a spectrum dominated by iron-group absorption, very different to the blue and largely featureless spectra of LFCs at peak.


Despite the difficulties with a pure nickel-powered model, white dwarf progenitor models (i.e. variants of SNe Ia) would still be appealing to explain transients in old stellar populations. A SN Ia interacting with a dense CSM could produce a peak luminosity well in excess of typical SNe Ia without requiring $M_{\rm Ni}\sim M_{\rm ej}$. However, our NTT observations at 70--80 days after explosion indicate that AT2022aedm had already faded below the luminosity of a SN Ia at the same epoch, making a hidden SN Ia unlikely. Moreover, known interacting SNe Ia produce spectra with strong, broad hydrogen emission and broad metal P Cygni lines \citep{Silverman2013}, very unlike the spectra of LFCs.

\subsubsection{Magnetar birth}


Rapidly rotating nascent magnetars are suspected to play a role in many luminous and/or rapid transients, such as SLSNe, gamma-ray bursts and fast radio bursts. Central engines are also thought to be required in the Cow-like RETs. Powering short-timescale, luminous events like LFCs would require a combination of rapid rotation to provide a large energy reservoir, a strong magnetic field to extract this energy quickly, and a low ejecta mass \citep[e.g.][]{Kasen2010}. We find that although the \textsc{mosfit} magnetar model \citep{Nicholl2017} is able to adequately fit the decline phase of AT2022aedm, it struggles to simultaneously match the fast rise, even with an ejecta mass $\lesssim 1$\,\M.

The environments of known LFCs are also a major problem for this model. SLSNe and long GRBs are thought to prefer low-metallicity, star-forming galaxies because these are favourable for rapidly-rotating core collapse. The massive, elliptical hosts of LFCs are decidedly unfavourable. Less than 1\% of core-collapse SNe occur in elliptical galaxies \citep{Irani2022a}, and {likely not all form magnetars. \citet{Kouveliotou1998} estimate that only $\sim10\%$ of core-collapses form magnetars, though more recent work finds a potentially larger but rather uncertain fraction of $0.4_{-0.28}^{+0.6}$ \citep{Beniamini2019}. In any case, it appears} unlikely that three unusual magnetar-forming explosions would all be found in elliptical galaxies. Moreover, the blue-shifted narrow H and He lines in the later spectra of AT2022aedm may indicate a dense wind pre-explosion, which would strip angular momentum and inhibit magnetar formation.

\subsubsection{Shock breakout}


Interaction with CSM is another mechanism thought to be responsible for many luminous or unusual transients, including (SL)SNe IIn, and fast events like SNe Ibn/Icn. Models for luminous SNe IIn generally invoke a massive CSM that releases energy slowly via post-shock diffusion \citep{Smith2007}. In the case of rapidly evolving transients, the more relevant models are shock breakout from an extended CSM or wind, which have been investigated by \citet{Chevalier2011} and \citet{Ginzburg2012}. This model provided a reasonable explanation for the rapid SN IIn, PTF09uj \citep{Ofek2010}, and other RETs \citep{Drout2014}. The early He~II emission in AT2022aedm is often seen during the shock breakout phase in normal Type II SNe \citep{Gal-Yam2014,Bruch2022}.

We apply the equations of \citet{Chevalier2011}, following the prescriptions from \citet{Margutti2014}, to estimate the ejecta and wind masses required in AT2022aedm. We set the input parameters based on our \textsc{superbol} results: peak time $t_{\rm peak}=8$\,days, total radiated energy $E_{\rm rad}=1.1\times10^{51}$\,erg, and breakout radius $R_{\rm bo}\approx2\times10^{15}$\,cm. The equations are degenerate in $M_{\rm ej}/E^2$, where $E$ is the kinetic energy of the explosion. We find $M_{\rm ej} = 0.02(E/10^{51}{\rm erg})^2$\,\M, with a wind density parameter\footnote{Defined as $D_*\equiv\rho r^2/(5\times10^{16}\,{\rm g\,cm}^{-1})$, where $\rho$ is the wind density at radius $r$.} $D_*=0.27$. For a wind velocity of 2700\,\kms, based on the blue-shifted absorption lines in the spectrum, this corresponds to a pre-explosion mass-loss rate $\dot{M}=0.7$\,\M\,yr$^{-1}$. 


\citet{Vinko2015} noted that Dougie could also be explained by a reasonable shock-powered model, with an estimated $\approx 8\times10^{50}$\,erg deposited in $\approx2.6$\,\M\ of CSM, but disfavoured this model based on the lack of shock-excited lines in the spectrum. AT2022aedm shows that emission lines can be very weak and short-lived in these events, perhaps making a CSM interpretation of Dougie more palatable. Nevertheless, the parameters we infer for AT2022aedm (if shock breakout is the dominant power source) are difficult to associate with any specific progenitor. For a standard $10^{51}$\,erg explosion, the low ejecta mass would be indicative of an ultra-stripped SN, yet the wind is H- and He-rich, and the unsustainable high mass-loss rate would require that it is lost in the years immediately before explosion. For a very energetic explosion with $10^{52}$\,erg, the implied ejecta mass is a more reasonable $\sim2$\,\M. But this large energy would likely require a massive progenitor, increasing the tension with the passive host galaxies of the LFCs.

\subsection{Stellar mass compact mergers}


The old stellar populations hosting AT2022aedm and other LFCs are more compatible with a compact object origin, rather than massive stars. However, we encountered inconsistencies interpreting these objects as white dwarf explosions or TDEs from massive BHs. Fortunately, recent years have seen rapid progress in understanding the diversity of transients resulting from mergers involving neutron stars (NS) and stellar mass BHs, and we compare to such models here.

Kilonovae, transients powered by the decay of heavy elements ejected from NS mergers, have been discovered in targeted follow-up of gravitational waves \citep{Abbott2017a,Margutti2021} and gamma-ray bursts \citep{Tanvir2013,Berger2013,Rastinejad2022}. However, LFCs are much brighter and bluer than any plausible kilonovae, in which low ejecta masses $<0.1$\,\M\ and large opacities conspire to produce faint, red transients visible for only a few days in the optical. While no definitive detections exist, kilonovae from NS-BH mergers are expected to be even fainter and redder than those from binary NSs \citep{Gompertz2023a}. 
AT2018kzr was a so-far unique event suggested to be the merger of a NS with a white dwarf \citep{McBrien2019,Gillanders2020}. It was somewhat brighter and longer lived than known kilonovae, peaking at $M\approx-18$\,mag. However, it still faded much faster than LFCs (Figure \ref{fig:lc}) and showed broad metal absorption lines resembling SNe Ic, inconsistent with our events (Figure \ref{fig:spec}).


\citet{Lyutikov2019} proposed that Cow-like RETs can arise from accretion-induced collapse following the merger of a carbon-oxygen white dwarf with an oxygen-neon-magnesium white dwarf. This model produces a magnetar in combination with a low ejecta mass. If one of the white dwarfs retained a surface hydrogen layer (type DA), a residual pre-collapse wind can help to explain the observed lines in the spectrum of AT2022aedm. This model may be a plausible progenitor channel for LFCs. However, it also predicts Cow-like non-thermal emission, which is ruled out in the case of AT2022aedm. This channel also has a short delay time, such that the host galaxies are likely to still be star-forming, in tension with the elliptical hosts of our objects. {\citet{Brooks2017} suggest another channel for RETs from the collapse of a long-lived remnant of a helium white dwarf merging with an oxygen-neon white dwarf, though these models were fainter than our LFCs. } 



Finally, mergers involving a main sequence or evolved star with a stellar mass BH have also been suggested as progenitors of Cow-like events. Given the older host environments in LFCs, mergers with Wolf-Rayet stars \citep{Metzger2022a} are probably disfavoured in this case. {A merger of a BH with a lower mass He core is more consistent with a long delay time. In this case the accretion rate is expected to be much lower, though additional luminosity could be produced by interaction of disk winds with stellar material lost earlier in the merger \citep{Metzger2017b}.}

\citet{Kremer2021,Kremer2023} presented models for tidal disruptions of main sequence stars by stellar mass BHs in dense clusters. Some of their wind-reprocessed models exhibit optical rise times and post-peak temperatures similar to LFCs. However, while bolometric luminosities can reach $\sim10^{44}$\,\ergs, most of this energy is emitted in the UV and X-ray regime, and none of the models reach the peak optical magnitudes of our objects. 

Noting the early similarity of AT2022aedm to some TDE spectra, we encourage a broader exploration of the parameter space for stellar-mass TDEs. Another important test of TDE models (stellar or intermediate mass) will be identifying globular clusters at the positions of LFCs. The \textit{James Webb Space Telescope} (JWST) can reach the peak of the globular cluster luminosity function ($\approx -7.5$\,mag; \citealt{Harris1979}) in 10\,ks with NIRCam for an event within 200\,Mpc. At the distance of AT2022aedm, it is possible to achieve the same constraint in $\approx100$\,ks. We note however that the angular extent of a typical globular cluster at this distance is comparable to the NIRCam pixel scale, making identification challenging.


\section{Conclusions}

We have presented a detailed analysis of AT2022aedm, a very unusual, rapidly evolving transient in a massive elliptical galaxy. It has a rise time $<10$ days, a luminous peak $M_g=-22$\,mag, and a fast decline of 2\,mag in the subsequent 15 days. The optical colours evolve quickly to the red during the decline phase, while the spectrum remains devoid of identifiable broad features throughout. We do however detect narrow emission lines of H$\alpha$ and He~II during the first few days, reminiscent of shock cooling in young SNe, and narrow blue-shifted absorption lines at later times.

We identify two previous transients, both with uncertain classifications, that share the key properties of bright peaks, fast declines, strong colour evolution and spectra without high-equivalent width emission or absorption lines. This suggests that AT2022aedm represents a well-observed example of a previously unrecognised class of luminous, fast-cooling events that we term LFCs. All three events occurred in passive galaxies, with offsets of $\sim4-10$\,kpc.

Their unique combination of properties poses challenges for any physical scenario. The passive environments disfavour a massive star origin, while the light curves and spectra are inconsistent with thermonuclear SNe interacting with a dense medium. Mergers of compact object binaries are unable to reproduce the peak luminosity. Tidal disruption of a star by an intermediate mass BH, as suggested for Dougie \citep{Vinko2015}, may struggle to produce the fast colour evolution. Instead we find that these events may be broadly consistent with an extreme shock breakout, though from an as-yet unknown progenitor. Alternatively, models of TDEs from stellar-mass BHs show promise, though current models emit too much of their luminosity in the UV and X-rays. Refinements of stellar-mass TDE models, and searches for dense local environments at the sites of LFCs, should help to confirm or rule out this scenario.


\vspace{2em}
We thank Brian Metzger, Eliot Quataert, and an anonymous referee for helful comments that improved this paper. We thank Anna Ho for sharing the data on AT2020bot. We also thank Joe Bright, Paul Scott and David Titterington for help with the AMI-LA observations. 
MN, AA and XS are supported by the European Research Council (ERC) under the European Union's Horizon 2020 research and innovation programme (grant agreement No.~948381) and by UK Space Agency Grant No.~ST/Y000692/1. PR is supported by STFC Grant 2742655. 
SJS acknowledges funding from STFC Grant ST/X006506/1 and ST/T000198/1. 
TP acknowledges the support by ANID through the Beca Doctorado Nacional 202221222222.
JV is supported by NKFIH-OTKA grant K-142534 from the National Research, Development and Innovation Office, Hungary.
T-W.C thanks the Max Planck Institute for Astrophysics for hosting her as a guest researcher.
MP is supported by a research grant (19054) from VILLUM FONDEN. 
PC acknowledges support via an Academy of Finland grant (340613; P.I. R.~Kotak).
PW acknowledges support from the Science and Technology Facilities Council (STFC) grant ST/R000506/1.
MF is supported by a Royal Society - Science Foundation Ireland University Research Fellowship.
MK is partially supported by the program Unidad de Excelencia María de Maeztu CEX2020-001058-M.
LG, CPG, MK and TEMB acknowledge support from Unidad de Excelencia Mar\'ia de Maeztu CEX2020-001058-M, from Centro Superior de Investigaciones Cient\'ificas (CSIC) under the PIE project 20215AT016, and from the Spanish Ministerio de Ciencia e Innovaci\'on (MCIN) and the Agencia Estatal de Investigaci\'on (AEI) 10.13039/501100011033 under the PID2020-115253GA-I00 HOSTFLOWS project. L.G. also acknowledges support from the European Social Fund (ESF) ``Investing in your future'' under the 2019 Ram\'on y Cajal program RYC2019-027683-I. CPG also acknowledges financial support from the Secretary of Universities and Research (Government of Catalonia) and by the Horizon 2020 Research and Innovation Programme of the European Union under the Marie Sklodowska-Curie and the Beatriu de Pin\'os 2021 BP 00168 programme. TEMB also acknowledges financial support from the 2021 Juan de la Cierva program FJC2021-047124-I. 
GPS acknowledges support from The Royal Society, the Leverhulme Trust, and the Science and Technology Facilities Council (grant numbers ST/N021702/1 and ST/S006141/1).
FEB acknowledges support from ANID-Chile BASAL CATA ACE210002 and FB210003, FONDECYT Regular 1200495,
and Millennium Science Initiative Program  – ICN12\_009.

ATLAS is primarily funded through NASA grants NN12AR55G, 80NSSC18K0284, and 80NSSC18K1575. The ATLAS science products are provided by the University of Hawaii, QUB, STScI, SAAO and Millennium Institute of Astrophysics in Chile. 
The Pan-STARRS telescopes are supported by NASA Grants NNX12AR65G and NNX14AM74G. 
Based on observations collected at the European Organisation for Astronomical Research in the Southern Hemisphere, Chile, as part of ePESSTO+ (the advanced Public ESO Spectroscopic Survey for Transient Objects Survey). ePESSTO + observations were obtained under ESO programme ID 108.220C (PI:Inserra). 
This work makes use of data from the Las Cumbres Observatory global network of telescopes. The LCO group is supported by NSF grants AST-1911151 and AST-1911225. 
The Liverpool Telescope is operated on the island of La Palma by Liverpool John Moores University in the Spanish Observatorio del Roque de los Muchachos of the Instituto de Astrofisica de Canarias with financial support from the UK Science and Technology Facilities Council. 
We thank the staff of the Mullard Radio Astronomy Observatory for their assistance in the maintenance and operation of AMI, which is supported by the Universities of Cambridge and Oxford.  We also acknowledge support from the European Research Council under grant ERC-2012-StG-307215 LODESTONE.


%

\vspace{5mm}
\facilities{NTT, PS1, Liverpool:2m, LCOGT, MMT, Swift, AMI}


\software{
Astropy \citep{Astropy2013,Astropy2018},
Matplotlib \citep{Hunter2007},
Numpy \citep{Harris2020},
SciPy \citep{Virtanen2020},
Astroquery \citep{Ginsburg2019},
Astrometry.net \citep{Lang2010},
Astroalign \citep{Beroiz2020},
Lacosmic \citep{vanDokkum2012},
Pyzogy \citep{Guevel2017},
Photutils \citep{Bradley2020},
Superbol \citep{Nicholl2018},
Mosfit \citep{Guillochon2018a},
Aladin \citep{Bonnarel2000},
PSF \citep{Nicholl2023}
          }





\bibliography{22aedm}{}
\bibliographystyle{aasjournal}



\end{document}